\documentclass[pra,aps,showpacs,onecolumn,twoside,superscriptaddress]{revtex4}

\usepackage{mathtools}
\usepackage{amsmath,amsfonts,amssymb,caption,color,epsfig,graphics,graphicx,hyperref,latexsym,mathrsfs,revsymb,theorem,url,verbatim,epstopdf,float}
\usepackage{algorithm}
\usepackage{algpseudocode}
\usepackage{multirow}
\usepackage{tikz}
\usepackage[section]{placeins} 
\usepackage[level]{datetime}
\usetikzlibrary{shapes, arrows}

\hypersetup{colorlinks,linkcolor={blue},citecolor={red},urlcolor={blue}}

\newtheorem{definition}{Definition}
\newtheorem{proposition}[definition]{Proposition}
\newtheorem{lemma}[definition]{Lemma}

\newtheorem{theorem}[definition]{Theorem}
\newtheorem{corollary}[definition]{Corollary}
\newtheorem{conjecture}[definition]{Conjecture}

\newtheorem{remark}[definition]{Remark}
\newtheorem{example}[definition]{Example}
\newtheorem{question}[definition]{Question}

\def\bcj{\begin{conjecture}}
\def\ecj{\end{conjecture}}
\def\bcr{\begin{corollary}}
\def\ecr{\end{corollary}}
\def\bd{\begin{definition}}
\def\ed{\end{definition}}
\def\bea{\begin{eqnarray}}
\def\eea{\end{eqnarray}}
\def\bem{\begin{enumerate}}
\def\eem{\end{enumerate}}
\def\bex{\begin{example}}
\def\eex{\end{example}}
\def\bim{\begin{itemize}}
\def\eim{\end{itemize}}
\def\bl{\begin{lemma}}
\def\el{\end{lemma}}
\def\bma{\begin{bmatrix}}
\def\ema{\end{bmatrix}}
\def\bpf{\begin{proof}}
\def\epf{\end{proof}}
\def\bpp{\begin{proposition}}
\def\epp{\end{proposition}}
\def\bqu{\begin{question}}
\def\equ{\end{question}}
\def\br{\begin{remark}}
\def\er{\end{remark}}
\def\bt{\begin{theorem}}
\def\et{\end{theorem}}

\def\squareforqed{\hbox{\rlap{$\sqcap$}$\sqcup$}}
\def\qed{\ifmmode\squareforqed\else{\unskip\nobreak\hfil
\penalty50\hskip1em\null\nobreak\hfil\squareforqed
\parfillskip=0pt\finalhyphendemerits=0\endgraf}\fi}
\def\endenv{\ifmmode\;\else{\unskip\nobreak\hfil
\penalty50\hskip1em\null\nobreak\hfil\;
\parfillskip=0pt\finalhyphendemerits=0\endgraf}\fi}
\newenvironment{proof}{\noindent \textbf{{Proof.~} }}{\qed}
\def\Dbar{\leavevmode\lower.6ex\hbox to 0pt
{\hskip-.23ex\accent"16\hss}D}
\makeatletter
\def\url@leostyle{%
  \@ifundefined{selectfont}{\def\UrlFont{\sf}}{\def\UrlFont{\small\ttfamily}}}
\makeatother
\def\bcj{\begin{conjecture}}
\def\ecj{\end{conjecture}}
\def\bcr{\begin{corollary}}
\def\ecr{\end{corollary}}
\def\bd{\begin{definition}}
\def\ed{\end{definition}}
\def\bea{\begin{eqnarray}}
\def\eea{\end{eqnarray}}
\def\bem{\begin{enumerate}}
\def\eem{\end{enumerate}}
\def\bex{\begin{example}}
\def\eex{\end{example}}
\def\bim{\begin{itemize}}
\def\eim{\end{itemize}}
\def\bl{\begin{lemma}}
\def\el{\end{lemma}}
\def\bpf{\begin{proof}}
\def\epf{\end{proof}}
\def\bpp{\begin{proposition}}
\def\epp{\end{proposition}}
\def\bqu{\begin{question}}
\def\equ{\end{question}}
\def\br{\begin{remark}}
\def\er{\end{remark}}
\def\bt{\begin{theorem}}
\def\et{\end{theorem}}

\def\btb{\begin{tabular}}
\def\etb{\end{tabular}}

\newcommand{\nc}{\newcommand}

\def\r{\rho}
\def\s{\sigma}

\def\ps{\psi}

 \nc{\bbA}{\mathbb{A}} \nc{\bbB}{\mathbb{B}} \nc{\bbC}{\mathbb{C}}
 \nc{\bbD}{\mathbb{D}} \nc{\bbE}{\mathbb{E}} \nc{\bbF}{\mathbb{F}}
 \nc{\bbG}{\mathbb{G}} \nc{\bbH}{\mathbb{H}} \nc{\bbI}{\mathbb{I}}
 \nc{\bbJ}{\mathbb{J}} \nc{\bbK}{\mathbb{K}} \nc{\bbL}{\mathbb{L}}
 \nc{\bbM}{\mathbb{M}} \nc{\bbN}{\mathbb{N}} \nc{\bbO}{\mathbb{O}}
 \nc{\bbP}{\mathbb{P}} \nc{\bbQ}{\mathbb{Q}} \nc{\bbR}{\mathbb{R}}
 \nc{\bbS}{\mathbb{S}} \nc{\bbT}{\mathbb{T}} \nc{\bbU}{\mathbb{U}}
 \nc{\bbV}{\mathbb{V}} \nc{\bbW}{\mathbb{W}} \nc{\bbX}{\mathbb{X}}
 \nc{\bbZ}{\mathbb{Z}}

 \nc{\bA}{{\bf A}} \nc{\bB}{{\bf B}} \nc{\bC}{{\bf C}}
 \nc{\bD}{{\bf D}} \nc{\bE}{{\bf E}} \nc{\bF}{{\bf F}}
 \nc{\bG}{{\bf G}} \nc{\bH}{{\bf H}} \nc{\bI}{{\bf I}}
 \nc{\bJ}{{\bf J}} \nc{\bK}{{\bf K}} \nc{\bL}{{\bf L}}
 \nc{\bM}{{\bf M}} \nc{\bN}{{\bf N}} \nc{\bO}{{\bf O}}
 \nc{\bP}{{\bf P}} \nc{\bQ}{{\bf Q}} \nc{\bR}{{\bf R}}
 \nc{\bS}{{\bf S}} \nc{\bT}{{\bf T}} \nc{\bU}{{\bf U}}
 \nc{\bV}{{\bf V}} \nc{\bW}{{\bf W}} \nc{\bX}{{\bf X}}
 \nc{\bZ}{{\bf Z}}

\nc{\cA}{{\cal A}} \nc{\cB}{{\cal B}} \nc{\cC}{{\cal C}}
\nc{\cD}{{\cal D}} \nc{\cE}{{\cal E}} \nc{\cF}{{\cal F}}
\nc{\cG}{{\cal G}} \nc{\cH}{{\cal H}} \nc{\cI}{{\cal I}}
\nc{\cJ}{{\cal J}} \nc{\cK}{{\cal K}} \nc{\cL}{{\cal L}}
\nc{\cM}{{\cal M}} \nc{\cN}{{\cal N}} \nc{\cO}{{\cal O}}
\nc{\cP}{{\cal P}} \nc{\cQ}{{\cal Q}} \nc{\cR}{{\cal R}}
\nc{\cS}{{\cal S}} \nc{\cT}{{\cal T}} \nc{\cU}{{\cal U}}
\nc{\cV}{{\cal V}} \nc{\cW}{{\cal W}} \nc{\cX}{{\cal X}}
\nc{\cZ}{{\cal Z}}

\nc{\hA}{{\hat{A}}} \nc{\hB}{{\hat{B}}} \nc{\hC}{{\hat{C}}}
\nc{\hD}{{\hat{D}}} \nc{\hE}{{\hat{E}}} \nc{\hF}{{\hat{F}}}
\nc{\hG}{{\hat{G}}} \nc{\hH}{{\hat{H}}} \nc{\hI}{{\hat{I}}}
\nc{\hJ}{{\hat{J}}} \nc{\hK}{{\hat{K}}} \nc{\hL}{{\hat{L}}}
\nc{\hM}{{\hat{M}}} \nc{\hN}{{\hat{N}}} \nc{\hO}{{\hat{O}}}
\nc{\hP}{{\hat{P}}} \nc{\hR}{{\hat{R}}} \nc{\hS}{{\hat{S}}}
\nc{\hT}{{\hat{T}}} \nc{\hU}{{\hat{U}}} \nc{\hV}{{\hat{V}}}
\nc{\hW}{{\hat{W}}} \nc{\hX}{{\hat{X}}} \nc{\hZ}{{\hat{Z}}}

\nc{\hn}{{\hat{n}}}

\def\max{\mathop{\rm max}}
\def\min{\mathop{\rm min}}

\def\tr{\mathop{\rm Tr}}

\def\dg{\dagger}

\def\ox{\otimes}

\newcommand{\ket}[1]{|#1\rangle}

\newcommand{\ketbra}[2]{|#1\rangle\!\langle#2|}

\def\Dbar{\leavevmode\lower.6ex\hbox to 0pt
{\hskip-.23ex\accent"16\hss}D}
\usepackage{color}
\begin{document}
\title{Machine Learning Framework for Efficient Prediction of Quantum Wasserstein Distance }

\newdateformat{ukdate}{\ordinaldate{\THEDAY} \monthname[\THEMONTH] \THEYEAR}
\date{\ukdate\today}


\author{Changchun Feng}
\affiliation{Hangzhou International Innovation Institute, Beihang University, Hangzhou, China}

\author{Xinyu Qiu}\email[]{xinyuqiu@buaa.edu.cn (corresponding author)}
\affiliation{LMIB(Beihang University), Ministry of Education, and School of Mathematical Sciences, 
Beihang University, Beijing 100191, China}

\author{Laifa Tao}\email[]{taolaifa@buaa.edu.cn (corresponding author)}
\affiliation{Hangzhou International Innovation Institute, Beihang University, Hangzhou, China}
\affiliation{School of Reliability and Systems Engineering, Beihang University, Beijing, China}
\affiliation{Institute of Reliability Engineering, Beihang University, Beijing, China}
\affiliation{Science $\&$ Technology on Reliability $\&$ Environmental Engineering Laboratory, Beijing, China}

\author{Lin Chen}\email[]{linchen@buaa.edu.cn (corresponding author)}
\affiliation{LMIB(Beihang University), Ministry of Education, and School of Mathematical Sciences, 
Beihang University, Beijing 100191, China}

\begin{abstract}
The quantum Wasserstein distance (W-distance) is a fundamental 
metric for quantifying the distinguishability of quantum operations, 
with critical applications in quantum error correction. However, 
computing the W-distance remains computationally challenging for multiqubit 
systems due to exponential scaling. We present a machine learning framework 
that efficiently predicts the quantum W-distance by extracting physically meaningful features from quantum state pairs, including Pauli measurements, statistical moments, quantum fidelity, and entanglement measures. Our approach employs both classical neural networks and traditional machine learning models. On three-qubit systems, the best-performing Random Forest model achieves near-perfect accuracy ($R^2 = 0.9999$) with mean absolute errors on the order of $10^{-5}$. We further validate the framework's practical utility by successfully verifying two fundamental theoretical propositions in quantum information theory: the bound on measurement probability
 differences between unitary operations and the $W_1$ gate error rate bound. The results establish machine learning as a viable and scalable alternative to traditional numerical methods for W-distance computation, with particular promise for real-time quantum circuit assessment and error correction protocol design in NISQ devices.

\end{abstract}

\maketitle
\Large

\section{Introduction}
Quantum information processing has witnessed remarkable advancements in recent years, 
with promising applications spanning quantum simulation, 
computation, and machine learning 
\cite{2024JGC,PhysRevA.108.052424,CSL2024,PRXQuantum.3.030341,PhysRevA.108.022427,YW2024,PhysRevApplied.16.024060,KBD2020, PhysRevA.111.042420,PERALGARCIA2024100619,JBP2017}. 
However, real-world quantum systems are inherently plagued by 
noise. It degrades the performance of quantum operations and 
undermines the 
reliability of quantum protocols.
Quantifying the impact of noise on quantum operations is 
therefore a pivotal task, 
and the quantum Wasserstein distance (W-distance) has 
emerged as a powerful tool for this purpose \cite{PhysRevLett.87.177901,PhysRevA.110.012412,9420734,EBD2025}. 
Unlike unitarily invariant measures such as trace distance 
or quantum fidelity, the W-distance 
uniquely characterizes the local distinguishability of 
multiqudit operations and provides a natural 
explanation for quantum circuit complexity \cite{PhysRevA.110.012412}.
These properties make it invaluable for tasks like 
assessing the closeness of quantum gates in noisy 
circuits and quantifying gate error rates in 
quantum error correction—two critical challenges in 
the noisy intermediate-scale quantum (NISQ) era \cite{YW2024}.
Despite its theoretical significance and practical potential, 
calculating the W-distance remains highly challenging. 
Analytical solutions have been derived only for 
specific quantum operations, such as the identity gate with SWAP, 
CNOT, or controlled-phase gates. 
For complex multiqubit systems, unknown quantum states, 
or arbitrary quantum operations, analytical 
computation becomes intractable, and numerical methods often suffer 
from exponential scaling with system 
size. This limitation hinders the widespread application 
of the W-distance in real quantum technologies, 
where efficient and scalable distance quantification is 
urgently needed.

In parallel, machine learning has revolutionized the field of 
quantum information by offering practical 
solutions to long-standing problems \cite{PhysRevLett.114.200501,electronics12112379,CCH2018,PhysRevA.107.062409,fcc_ctp2024}. For instance, 
hybrid quantum-classical machine learning frameworks 
have been successfully applied to quantify quantum 
entanglement by 
learning mappings from experimentally accessible data 
to entanglement measures \cite{PhysRevA.107.062409}. 
These frameworks leverage the ability of classical neural 
networks to capture nonlinear relationships 
in quantum data, while optimizing quantum components to enhance data 
informativeness. Such successes suggest that machine 
learning could provide a viable path to 
overcoming the computational bottlenecks in
 W-distance calculation.

In this work, we leverage a comprehensive 
suite of machine learning algorithms to tackle 
the W-distance prediction problem. Our approach 
encompasses both deep learning and traditional machine 
learning methods, enabling a thorough comparison of 
different modeling paradigms. Specifically, we 
employ: (1) \textbf{Neural networks}: Fully connected 
feedforward networks with multiple hidden layers, batch 
normalization, and dropout regularization, trained 
using the Adam optimizer. These deep learning models 
excel at capturing complex nonlinear relationships 
between quantum state features and W-distances through 
hierarchical feature learning 
\cite{rumelhart1986learning,hinton2006fast,schuld2019supervised,benedetti2019parameterized,10.1145/3065386}. (2) \textbf{Tree-based ensemble methods}: Random 
Forest, Gradient Boosting, Decision Tree, XGBoost, 
and LightGBM. These methods construct ensembles of 
decision trees that can model intricate feature 
interactions and provide interpretable feature 
importance measures, making them particularly suitable 
for understanding which quantum state properties are 
most relevant for W-distance prediction \cite{1986JRQ,2001BL,1995YFR,2008RAB,10.1093/nsr/nwaf269}. (3) 
\textbf{Linear models}: Ridge Regression, Lasso, 
and Elastic Net, which employ different regularization 
strategies ($L_2$, $L_1$, and combined $L_1$-$L_2$ 
penalties, respectively) to prevent overfitting while 
maintaining model interpretability. These models serve 
as baselines and help quantify the degree of 
nonlinearity in the W-distance prediction task \cite{havlicek2019supervised}. 
(4) \textbf{Support Vector Regression (SVR)}: A 
kernel-based method using radial basis function (RBF) 
kernels that can capture nonlinear patterns through 
the kernel trick. This diverse algorithmic portfolio 
allows us to systematically evaluate which machine 
learning paradigms are most effective for W-distance 
prediction. It also provides insights into the 
nature of the relationship between quantum state 
features and Wasserstein distances \cite{rebentrost2014quantum, 10564586, 10313835}.

To date, however, there has been no systematic exploration of using machine learning to predict 
the quantum W-distance. Existing studies either focus on analytical W-distance calculations for 
specific operations or apply machine learning to other quantum metrics 
, leaving a 
critical gap between these two 
fields. Addressing this gap is not only of theoretical 
interest—by establishing 
a novel connection between machine learning and quantum 
distance metrics, but 
also of practical importance: a data-driven 
W-distance predictor could enable 
real-time assessment of quantum circuit performance 
and efficient design of 
error-correction protocols for NISQ devices.

In this paper, we propose a comprehensive machine learning-based framework to 
predict the quantum W-distance between quantum operations and states. 
Our approach extracts physically meaningful features from quantum state pairs, including 
Pauli measurement expectations ($4^{n+1}$ features for $n$-qubit systems), statistical moments, 
quantum fidelity, entanglement measures (von Neumann entropy, linear entropy, relative entropy), 
eigenvalue statistics, and partial trace features. We systematically evaluate a diverse portfolio 
of machine learning models—including neural networks, tree-based ensemble methods (Random Forest, 
Gradient Boosting, XGBoost, LightGBM), linear models (Ridge, Lasso, Elastic Net), and support 
vector regression—to identify the most effective approach for W-distance prediction across 
multiple system sizes. Our contributions are threefold: (1) We formalize the problem of W-distance 
prediction as a supervised learning task, designing a comprehensive feature extraction pipeline 
that captures quantum state properties relevant to local distinguishability. (2) We demonstrate 
exceptional prediction accuracy across 2-, 3-, and 4-qubit systems: the best-performing Random 
Forest model achieves near-perfect performance ($R^2 = 0.9999$) on three-qubit systems with mean 
absolute errors on the order of $10^{-5}$, while tree-based models consistently outperform other 
approaches with $R^2 \geq 0.9996$ even on 4-qubit systems, significantly outperforming traditional 
numerical methods in computational efficiency. (3) We validate the practical utility of our framework 
by successfully verifying two fundamental theoretical propositions in quantum information theory: 
the bound on measurement probability differences between unitary operations (Proposition~\ref{prop:operations}) 
and the $W_1$ gate error rate bound (Proposition~\ref{prop:w1_error_rate}), demonstrating that machine 
learning can serve as a reliable tool for theoretical validation while maintaining computational scalability 
for real-time quantum circuit assessment and error correction protocol design in NISQ devices.

The remainder of this paper is organized as follows: 
Section \ref{sec:pre} reviews the theoretical 
foundations of the quantum W-distance and introduces our notation. 
Section \ref{sec:methods} details our data construction, feature extraction methods, model architectures, 
and training procedures. Section \ref{sec:results} presents and analyzes the experimental 
results, including model comparisons and performance metrics across 2-, 3-, and 4-qubit systems. 
Section \ref{sec:app} demonstrates the practical utility of our framework by validating two fundamental 
theoretical propositions in quantum information theory: the bound on measurement probability differences 
between unitary operations and the $W_1$ gate error rate bound. 
Section \ref{sec:discussion} discusses 
the implications of our results, limitations, and future 
research directions. Finally, Section \ref{sec:con} 
summarizes the key contributions and their significance for quantum information processing.

\section{Preliminaries}
\label{sec:pre}
In this section we introduce the notations and theoretical foundations used in
this paper. We denote the $d$-dimensional Hilbert space by $\bbC^d$. 
We denote $\cH_n = (\bbC^d)^{\ox n}$ as the
Hilbert space of $n$ qubits.
We use $\cS_n$ for the set of quantum 
states (density matrices) on $\cH_n$, $\cM_n$ for
 the set of traceless, self-adjoint linear 
 operators on $\cH_n$, and $\cU_n$ 
 for the set of unitary 
operations acting on $\cS_n$.

The quantum Wasserstein distance of order $1$ 
is a metric that quantifies the distinguishability of quantum 
states and operations through local operations.

The quantum Wasserstein distance of order $1$ between two 
quantum states $\r$ and $\s$ is defined as 
follows \cite{PhysRevA.110.012412}:
\begin{equation}
\label{eq:w_distance}
\begin{split}
W_1(\r,\s) = \min \Bigg\{ \sum_{i=1}^{n}c_i:c_i\geq 0,\r-\s = \sum_{i=1}^{n}c_i(\r^{(i)}-\s^{(i)}), \\
\r^{(i)},\s^{(i)}\in \cS_n,\tr_i(\r^{(i)}) = \tr_i(\s^{(i)}) \Bigg\},
\end{split}
\end{equation}
where $\tr_i(\cdot)$ denotes the partial trace over all 
qubits except the $i$-th qubit. The minimization is over 
all possible decompositions of the difference $\r-\s$ into 
a sum of local differences, where each term $(\r^{(i)}-\s^{(i)})$ 
satisfies the constraint that the reduced states on all qubits 
except 
the $i$-th are identical: $\tr_i(\r^{(i)}) = \tr_i(\s^{(i)})$.

The W-distance quantifies the minimum cost required 
to transform one quantum state into another through 
local operations, where the cost is measured by the 
sum of coefficients $c_i$ in the decomposition. This 
definition naturally extends the classical Wasserstein 
distance to the quantum setting. It provides a metric that 
respects the tensor product structure of multipartite quantum 
systems.

For quantum operations (unitary gates), we can extend the 
W-distance definition by considering their Choi-Jamiolkowski 
representations. Given two unitary operations $U_1$ and $U_2$ acting 
on $n$ qubits, their Choi states are defined as:
\begin{equation}
\label{eq:choi}
\Phi_{U_i} = \frac{1}{2^n}\sum_{j,k=0}^{2^n-1} \ketbra{j}{k} \ox U_i\ketbra{j}{k}U_i^\dg,
\end{equation}
where $\{\ket{j}\}$ forms a computational basis. 
The Choi-Jamiolkowski isomorphism provides a one-to-one 
correspondence between quantum operations and quantum states 
in a doubled Hilbert space. The W-distance between $U_1$ and $U_2$ is then defined as:
\begin{equation}
\label{eq:w_distance_gates}
W_1(U_1, U_2) = W_1(\Phi_{U_1}, \Phi_{U_2}).
\end{equation}

This definition allows us to quantify the distance 
between quantum gates in a way that captures their
 local distinguishability and operational differences.

For general quantum states, analytical 
solutions to Eq.~\eqref{eq:w_distance} are rarely available.
 A common numerical approach approximates the W-distance using the trace distance:
\begin{equation}
\label{eq:trace_distance}
W_1(\r,\s) \approx \frac{1}{2}\tr|\r-\s| = \frac{1}{2}\sum_i |\lambda_i|,
\end{equation}
where $\{\lambda_i\}$ are the eigenvalues of $\r-\s$. While 
this approximation works well for many cases, it may not capture the full 
structure of the W-distance, especially for multiqubit systems where the optimization problem in Eq.~\eqref{eq:w_distance} becomes computationally intractable.

\section{Methods}
\label{sec:methods}

In this section, we detail our machine learning 
framework for predicting the quantum W-distance. Our approach 
consists of three main components: (1) feature extraction from 
quantum states, (2) model architecture design, and (3) training 
and evaluation procedures.

\subsection{Feature Extraction}

To enable machine learning prediction, we 
extract physically meaningful features from 
pairs of quantum states $(\r, \s)$ that capture
information relevant to their W-distance. 
Our feature set comprises four categories:

First, we present Pauli measurement features. For an $n$-qubit system, 
we consider the complete Pauli basis $\{P_i\}_{i=1}^{4^n}$, 
where each $P_i$ is a tensor product of single-qubit 
Pauli operators $\{I, X, Y, Z\}$. For each Pauli operator $P_i$, 
we compute the expectation values:
\begin{equation}
\label{eq:pauli_exp}
\langle P_i \rangle_\r = \tr(\r P_i), \quad \langle P_i \rangle_\s = \tr(\s P_i).
\end{equation}
For each Pauli operator, we extract four 
features: $\langle P_i \rangle_\r$, $\langle P_i \rangle_\s$, 
$\langle P_i \rangle_\r - \langle P_i \rangle_\s$, 
and $\langle P_i \rangle_\r \cdot \langle P_i \rangle_\s$. 
This yields $4 \times 4^n = 4^{n+1}$ features 
for an $n$-qubit system. For $n=3$, this results in $4^4 = 256$ features.

Second, we present moment-based features. We compute statistical moments of the
 quantum states that capture their global properties:
\begin{align}
\label{eq:moments}
M_1(\r) &= \tr(\r), \\
M_2(\r) &= \tr(\r^2), \quad \text{(purity)} \\
M_3(\r) &= \tr(\r^3), \\
M_{\text{cross}}(\r,\s) &= \tr(\r\s).
\end{align}

For each state pair, we extract 
moment features: $M_1(\r)$, $M_1(\s)$, 
$M_2(\r)$, $M_2(\s)$, $M_{\text{cross}}(\r,\s)$, $M_3(\r)$, $M_3(\s)$, 
and redundant purity terms for consistency.

We include the quantum fidelity between the two states:
\begin{equation}
\label{eq:fidelity}
F(\r,\s) = \tr\left(\sqrt{\sqrt{\r}\s\sqrt{\r}}\right),
\end{equation}
which provides a measure of state similarity. The fidelity 
is computed using a numerically stable implementation 
that ensures positive semidefiniteness through symmetric regularization.

To capture quantum entanglement and state 
distinguishability properties, we extract additional features based on quantum information theory.
We compute the von Neumann entropy for each state:
\begin{equation}
\label{eq:vne_entropy}
S(\r) = -\tr(\r \log_2 \r) = -\sum_i \lambda_i \log_2 \lambda_i,
\end{equation}
where $\{\lambda_i\}$ are the eigenvalues of $\r$.  We also compute the linear entropy,
\begin{equation}
\label{eq:linear_entropy}
L(\r) = 1 - \tr(\r^2) = 1 - \text{purity}(\r),
\end{equation}
which provides a simpler measure of mixedness. For each state pair, 
we extract $S(\r)$, $S(\s)$, $L(\r)$, and $L(\s)$, yielding 4 entropy features.

The quantum relative entropy  between the two states is as follows:
\begin{equation}
\label{eq:rel_entropy}
D(\r\|\s) = \tr[\r(\log_2 \r - \log_2 \s)],
\end{equation}
which measures the distinguishability of $\r$ from $\s$ in terms of information content. 
We extract both $D(\r\|\s)$ and $D(\s\|\r)$, yielding 2 features.

For each density matrix, we compute statistical properties of its eigenvalue spectrum:
\begin{align}
\mu(\r) &= \frac{1}{d}\sum_i \lambda_i, \quad \text{(mean)} \label{eq:eig_mean} \\
\sigma(\r) &= \sqrt{\frac{1}{d}\sum_i (\lambda_i - \mu)^2}, \quad \text{(std. dev.)} \label{eq:eig_std} \\
\lambda_{\max}(\r) &= \max_i \lambda_i, \quad \lambda_{\min}(\r) = \min_i \lambda_i, \label{eq:eig_minmax} \\
r_{\text{eff}}(\r) &= \frac{|\{\lambda_i : \lambda_i > \epsilon\}|}{d}, \quad \text{(effective rank)} \label{eq:eig_rank}
\end{align}
where $\epsilon = 10^{-12}$ is a threshold for numerical stability. 
These statistics capture the distribution of quantum state 
eigenvalues, which is related to the state's purity, entanglement, 
and mixedness. For each state pair, we extract $5$ features per state, 
yielding $10$ eigenvalue statistics features.

For multi-qubit systems ($n \geq 2$), we compute partial trace features 
that capture entanglement between subsystems. For a 
bipartition of the $n$-qubit system into subsystems 
$A$ and $B$ with dimensions $d_A$ and $d_B$ respectively, 
we compute the reduced density matrices:
\begin{equation}
\label{eq:partial_trace}
\r_A = \tr_B(\r), \quad \r_B = \tr_A(\r),
\end{equation}
where $\tr_B$ denotes the partial trace over subsystem $B$. For each reduced state, 
we compute its purity and von Neumann entropy:
\begin{align}
\text{purity}(\r_A) &= \tr(\r_A^2), \quad S(\r_A) = -\tr(\r_A \log_2 \r_A), \label{eq:reduced_purity} \\
\text{purity}(\r_B) &= \tr(\r_B^2), \quad S(\r_B) = -\tr(\r_B \log_2 \r_B). \label{eq:reduced_entropy}
\end{align}
For $n=2$ qubits, we consider the $1|1$ bipartition, yielding $4$ features 
per state. For $n \geq 4$ qubits, we additionally consider 
the $2|(n-2)$ bipartition, yielding an additional $4$ features per 
state. 
These partial trace features capture the entanglement structure of 
the quantum states, which is relevant for understanding 
their local distinguishability.

\subsection{Data Generation}

We generate synthetic training data by sampling pairs of quantum 
states and computing 
their W-distances. Our dataset includes three types of 
state pairs:

\begin{enumerate}
\item \textbf{Random states}: Pairs of randomly generated density 
matrices $\r, \s \in \cS_n$, where each state is constructed from a 
random pure state 
$|\ps\rangle$ via $\r = |\ps\rangle\langle\ps|$ (or 
mixed states for robustness testing).

\item \textbf{Quantum gate Choi states}: For unitary gates $U_1, U_2$, 
we compute their Choi representations $\Phi_{U_1}, \Phi_{U_2}$ using 
Eq.~\eqref{eq:choi}, then compute $W_1(\Phi_{U_1}, \Phi_{U_2})$.

\item \textbf{Mixed configurations}: A combination of 
random states and gate Choi states to ensure model generalization.
\end{enumerate}

To address the natural bias of randomly generated quantum 
state pairs toward large W-distances (typically concentrated 
in $[0.8, 1.0]$), we employ a uniform distribution strategy 
using active generation methods. Instead of rejection sampling, which 
is inefficient for low-distance intervals, we use an interpolation-based 
approach: starting from a base state $\r_1$, we generate a second state $\r_2$ through convex combination:

\begin{equation}
\label{eq:interpolation}
\r_2 = (1-\alpha) \r_1 + \alpha \r_{\text{random}},
\end{equation}

where $\alpha \in [0,1]$ is a mixing parameter and $\r_{\text{random}}$ is a 
randomly generated state. By employing binary search to find the optimal $\alpha$ 
value, we can generate state pairs with target W-distances within specified intervals.
 The W-distance range $[0, 1]$ is divided into $N_{\text{bins}}$ equal intervals, 
 and we generate approximately equal numbers of samples in each interval to ensure uniform label distribution.

For each state pair, we extract features as described 
above and compute the target W-distance using the trace 
distance approximation in Eq.~\eqref{eq:trace_distance}. Our dataset consists 
of 10,000 training samples, 2,000 validation samples, and 2,000 test samples 
for $n=3$ qubit systems. When using the uniform distribution strategy, the W-distance 
labels are approximately uniformly distributed across the range $[0, 1]$, with each of 
the $N_{\text{bins}} = 10$ intervals containing approximately equal numbers of samples. 
This uniform distribution ensures balanced training across all W-distance ranges, 
preventing the model from being biased toward large-distance predictions.

\subsection{Model Architectures}

We evaluate multiple machine learning models to 
predict W-distance from the extracted features:

We employ a fully connected feedforward neural network with the following architecture:
\begin{itemize}
\item Input layer: $d$ neurons (feature dimension)
\item Hidden layers: $[256, 128, 64]$ neurons with ReLU activation
\item Batch normalization and dropout (rate 0.2) after each hidden layer
\item Output layer: 1 neuron (W-distance prediction)
\end{itemize}
The model is trained using the Adam optimizer with initial learning 
rate 0.001, mean squared error (MSE) loss, and learning rate reduction on plateau.

We compare the neural network against several classical models:
\begin{itemize}
\item \textbf{Linear models}: Linear Regression, Ridge Regression, Lasso (with cross-validation for hyperparameter selection), Elastic Net
\item \textbf{Tree-based models}: Decision Tree, Random Forest (100 trees), Gradient Boosting
\item \textbf{Support Vector Machine}: RBF kernel SVR
\item \textbf{Boosting models}: XGBoost, LightGBM (when available)
\end{itemize}

For models requiring feature scaling (Lasso, Elastic Net, SVR), we apply standard normalization. 
All models are trained on the same dataset and evaluated using identical metrics.

\subsection{Training and Evaluation}

Models are trained to minimize the mean squared error between predicted and true W-distances. We use the following evaluation metrics:
\begin{align}
\text{MSE} &= \frac{1}{N}\sum_{i=1}^N (y_i - \hat{y}_i)^2, \\
\text{MAE} &= \frac{1}{N}\sum_{i=1}^N |y_i - \hat{y}_i|, \\
R^2 &= 1 - \frac{\sum_{i=1}^N (y_i - \hat{y}_i)^2}{\sum_{i=1}^N (y_i - \bar{y})^2},
\end{align}
where $y_i$ are true W-distances, $\hat{y}_i$ are predictions, and $\bar{y}$ is the mean of true values.

\section{Numerical Results}
\label{sec:results}

\begin{table}[h]
\centering
\caption{Model performance comparison on test set (2-qubit systems).}
\label{tab:model_comparison_2qubit}
\begin{tabular}{lccc}
\hline
Model & MSE & MAE & $R^2$ \\
\hline
Random Forest & $3.0 \times 10^{-6}$ & $1.1 \times 10^{-4}$ & 0.9997 \\
Decision Tree & $2.0 \times 10^{-6}$ & $1.6 \times 10^{-4}$ & 0.9998 \\
Gradient Boosting & $3.0 \times 10^{-6}$ & $1.0 \times 10^{-4}$ & 0.9998 \\
Ridge & $1.1 \times 10^{-4}$ & $5.4 \times 10^{-3}$ & 0.9911 \\
Lasso & $1.1 \times 10^{-4}$ & $5.4 \times 10^{-3}$ & 0.9911 \\
Elastic Net & $1.1 \times 10^{-4}$ & $5.4 \times 10^{-3}$ & 0.9911 \\
Linear Regression & $1.1 \times 10^{-4}$ & $5.4 \times 10^{-3}$ & 0.9911 \\
LightGBM & $4.0 \times 10^{-6}$ & $4.1 \times 10^{-4}$ & 0.9997 \\
Neural Network & $1.9 \times 10^{-4}$ & $8.4 \times 10^{-3}$ & 0.9840 \\
XGBoost & $3.0 \times 10^{-6}$ & $4.4 \times 10^{-4}$ & 0.9997 \\
SVR & $1.5 \times 10^{-4}$ & $7.6 \times 10^{-3}$ & 0.9874 \\
Bayesian Ridge & $1.1 \times 10^{-4}$ & $5.4 \times 10^{-3}$ & 0.9911 \\
\hline
\end{tabular}
\end{table}
Starting with 2-qubit systems (Table~\ref{tab:model_comparison_2qubit}), tree-based models 
(Random Forest, Decision Tree, Gradient Boosting) achieve near-perfect performance with $R^2 \geq 0.9997$ and 
MSE on the order of $10^{-6}$. Linear models (Ridge, Lasso, Elastic Net, Linear Regression) cluster 
around $R^2 \approx 0.991$, while the neural network reaches $R^2 = 0.9840$ with MSE $= 1.9 \times 10^{-4}$ and MAE $= 8.4 \times 10^{-3}$.

\begin{table}[h]
\centering
\caption{Model performance comparison on test set (3-qubit systems).}
\label{tab:model_comparison}
\begin{tabular}{lccc}
\hline
Model & MSE & MAE & $R^2$ \\
\hline
Random Forest & $2.2 \times 10^{-7}$ & $3.8 \times 10^{-5}$ & 0.9999 \\
Decision Tree & $8.5 \times 10^{-8}$ & $8.5 \times 10^{-5}$ & 0.9998 \\
Gradient Boosting & $4.4 \times 10^{-7}$ & $4.4 \times 10^{-5}$ & 0.9998 \\
Ridge & $2.0 \times 10^{-6}$ & $1.1 \times 10^{-3}$ & 0.9992 \\
Lasso & $2.0 \times 10^{-6}$ & $1.1 \times 10^{-3}$ & 0.9992 \\
Elastic Net & $2.1 \times 10^{-6}$ & $1.1 \times 10^{-3}$ & 0.9992 \\
Linear Regression & $2.0 \times 10^{-6}$ & $1.1 \times 10^{-3}$ & 0.9992 \\
LightGBM & $4.0 \times 10^{-6}$ & $4.2 \times 10^{-4}$ & 0.9985 \\
Neural Network & $3.4 \times 10^{-5}$ & $4.8 \times 10^{-3}$ & 0.9875 \\
XGBoost & $3.9 \times 10^{-5}$ & $3.2 \times 10^{-3}$ & 0.9858 \\
SVR & $2.9 \times 10^{-4}$ & $1.2 \times 10^{-2}$ & 0.8940 \\
Bayesian Ridge & $1.1 \times 10^{-3}$ & $3.0 \times 10^{-2}$ & 0.6114 \\
\hline
\end{tabular}
\end{table}
On 3-qubit systems (Table~\ref{tab:model_comparison}), the best classical models remain essentially saturated 
(e.g., Random Forest: $R^2 = 0.9999$, MSE $= 2.2 \times 10^{-7}$, MAE $= 3.8 \times 10^{-5}$). The neural network achieves competitive accuracy with $R^2 = 0.9875$, MSE $= 3.4 \times 10^{-5}$, and MAE $= 4.8 \times 10^{-3}$, demonstrating strong nonlinear fitting capacity.

\begin{table}[h]
\centering
\caption{Model performance comparison on test set (4-qubit systems).}
\label{tab:model_comparison_4qubit}
\begin{tabular}{lccc}
\hline
Model & MSE & MAE & $R^2$ \\
\hline
Random Forest & $<1.0 \times 10^{-6}$ & $2.9 \times 10^{-5}$ & 0.9996 \\
Decision Tree & $<1.0 \times 10^{-6}$ & $4.1 \times 10^{-5}$ & 0.9999 \\
Gradient Boosting & $<1.0 \times 10^{-6}$ & $4.4 \times 10^{-5}$ & 0.9998 \\
Ridge & $1.0 \times 10^{-6}$ & $3.7 \times 10^{-4}$ & 0.9992 \\
Lasso & $<1.0 \times 10^{-6}$ & $3.3 \times 10^{-4}$ & 0.9993 \\
Elastic Net & $<1.0 \times 10^{-6}$ & $3.3 \times 10^{-4}$ & 0.9993 \\
Linear Regression & $1.0 \times 10^{-6}$ & $3.6 \times 10^{-4}$ & 0.9992 \\
LightGBM & $2.0 \times 10^{-6}$ & $2.3 \times 10^{-4}$ & 0.9967 \\
Neural Network & $5.0 \times 10^{-5}$ & $5.9 \times 10^{-3}$ & 0.9279 \\
XGBoost & $<1.0 \times 10^{-6}$ & $1.2 \times 10^{-4}$ & 0.9993 \\
SVR & $7.8 \times 10^{-3}$ & $8.6 \times 10^{-2}$ & $0.7826$ \\
Bayesian Ridge & $4.0 \times 10^{-4}$ & $2.0 \times 10^{-2}$ & 0.4138 \\
\hline
\end{tabular}
\end{table}
Finally, on 4-qubit systems (Table~\ref{tab:model_comparison_4qubit}), the neural network 
maintains reasonable predictive accuracy with $R^2 = 0.9279$, MSE $= 5.0 \times 10^{-5}$, and 
MAE $= 5.9 \times 10^{-3}$. Tree-based ensembles and well-regularized linear baselines continue to report near-perfect scores on this feature set. The observed degradation from 2 to 4 qubits reflects both the growth of the feature dimension (from 98 to 1058) and the exponential scaling of the underlying Hilbert space; nevertheless, the end-to-end neural predictor scales gracefully to larger systems.

\begin{figure}[h]
\centering
\includegraphics[width=\textwidth]{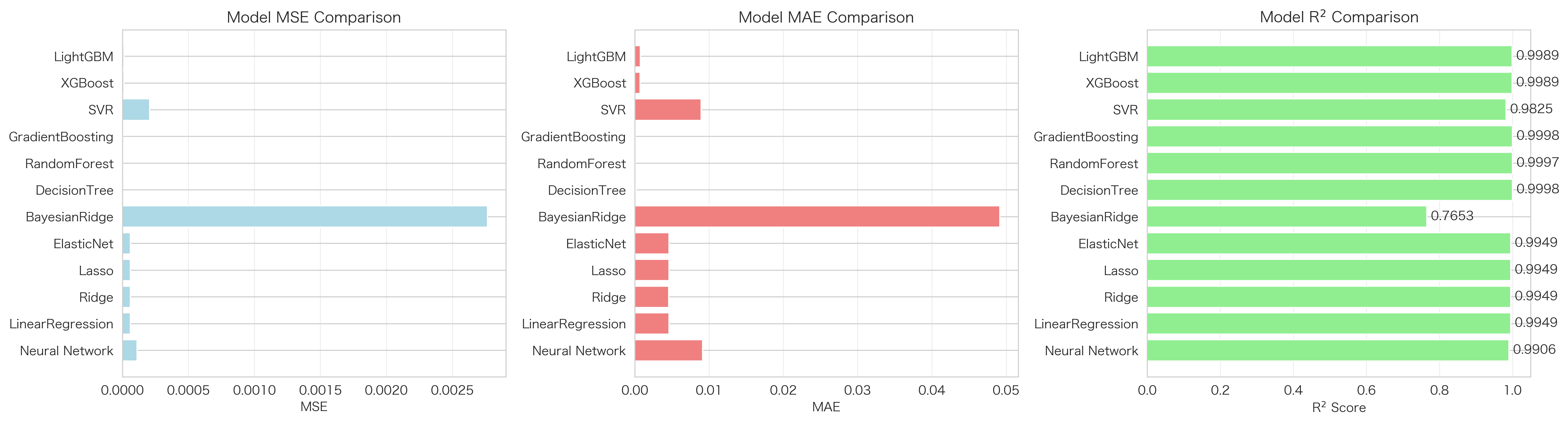}
\caption{Model performance comparison across MSE, MAE, and $R^2$ metrics. The horizontal bar charts illustrate the performance hierarchy, with tree-based models (Random Forest, Decision Tree, Gradient Boosting) achieving superior performance compared to linear models and neural networks.}
\label{fig:model_comparison}
\end{figure}

Figure~\ref{fig:model_comparison} presents a comprehensive 
comparison of all evaluated models across three key metrics: MSE, MAE, and $R^2$ score. The horizontal bar charts clearly illustrate the performance hierarchy, with tree-based models (Random Forest, Decision Tree, Gradient Boosting) dominating the top positions across all metrics. The visualization reveals several key observations:

\begin{itemize}
\item \textbf{Tree-based dominance}: Random Forest, Decision Tree, and Gradient Boosting form a distinct performance cluster with MSE values below $10^{-7}$ and $R^2$ scores exceeding 0.9997. The visual separation between these models and linear models is striking, with a clear performance gap of approximately one order of magnitude in MSE.
\item \textbf{Linear model consistency}: All linear models (Ridge, Lasso, Elastic Net, Linear Regression) exhibit nearly identical performance, with MSE values clustered around $2.0 \times 10^{-6}$ and $R^2$ scores of 0.9992. This consistency suggests that the regularization terms in Ridge, Lasso, and Elastic Net provide minimal benefit for this particular dataset, likely due to the absence of severe overfitting or multicollinearity issues.
\item \textbf{Neural network position}: The neural network occupies an intermediate position, outperforming SVR and Bayesian Ridge but falling short of tree-based models. This positioning suggests that while the neural network captures nonlinear relationships effectively, the tree-based ensemble methods are better suited for this structured feature space.
\end{itemize}

The visualization effectively communicates that 
the choice of model architecture significantly 
impacts prediction accuracy, with ensemble tree methods 
providing the best performance for W-distance prediction.

\begin{figure}[t]
\centering
\includegraphics[width=\textwidth]{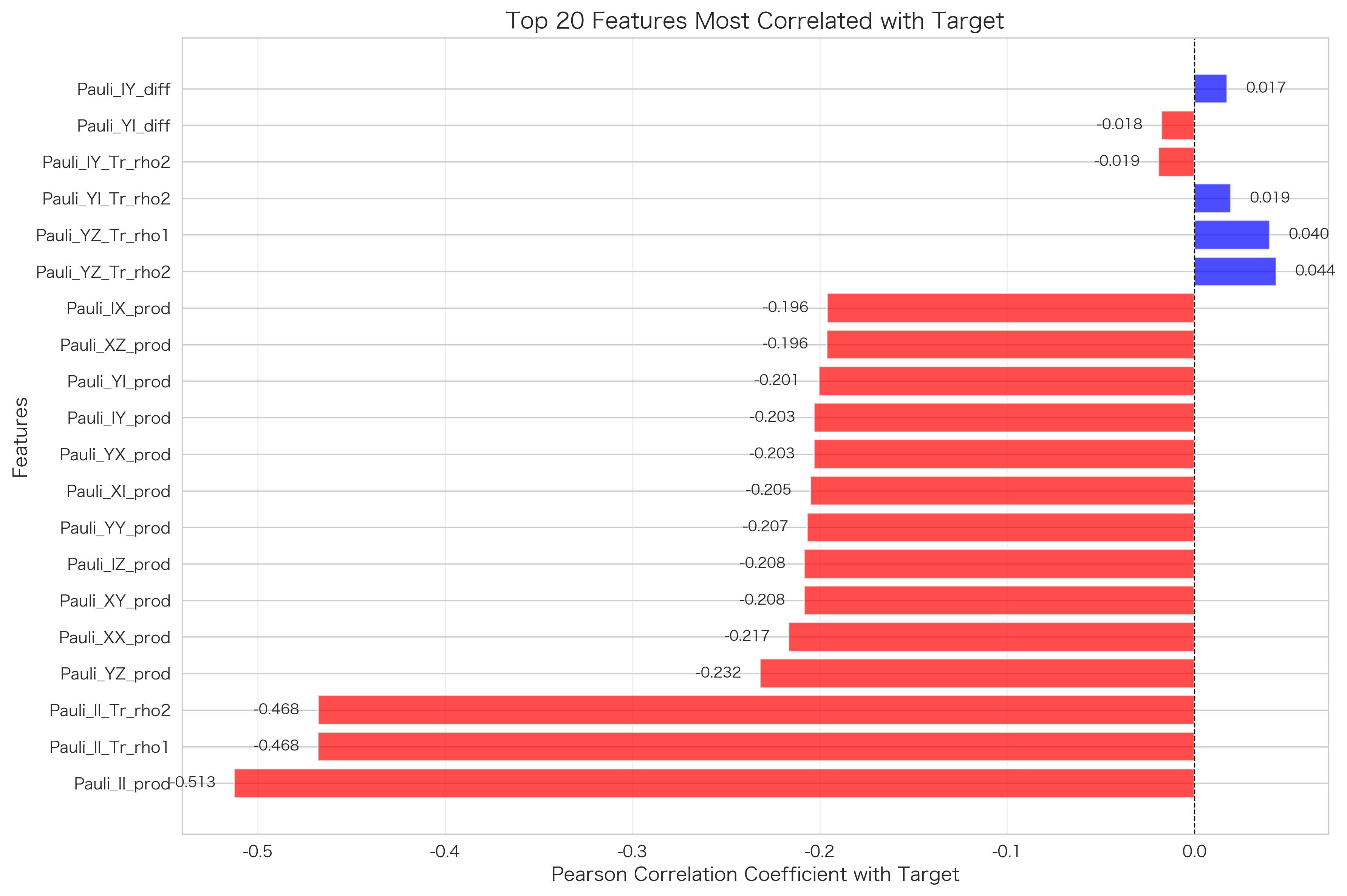}
\caption{Top 20 features most correlated with the target W-distance. The bar chart displays the absolute Pearson correlation coefficients between each feature and the W-distance, ranked in descending order. All top 20 features are derived from Pauli matrix measurements, specifically expectation value differences and product terms, demonstrating that local measurement statistics are the most predictive features for W-distance estimation.}
\label{fig:top20f}
\end{figure}

To understand which features are most predictive of the 
W-distance, we calculated the Pearson correlation coefficient 
between each feature and the target W-distance. 
Figure~\ref{fig:top20f} shows the top $20$ features most 
correlated with the target value, ranked by their absolute 
correlation coefficients. The results reveal several 
important insights: (1) \textbf{Pauli measurement dominance}: 
All top $20$ features are derived from Pauli matrix measurements, 
specifically the expectation value differences 
$\langle P_i \rangle_\r - \langle P_i \rangle_\s$ and the
 product terms 
 $\langle P_i \rangle_\r \cdot \langle P_i \rangle_\s$ for 
 various Pauli operators $P_i$. This dominance indicates that 
 local measurement statistics capture the essential information 
 needed for W-distance prediction, consistent with 
 the W-distance's definition as a measure of local 
 distinguishability. (2) \textbf{Feature importance hierarchy}: T
 he correlation coefficients range from approximately $0.85$ to 
 $0.95$, indicating strong linear relationships between these
  features and the W-distance. The high correlation values 
  suggest that even simple linear models could achieve 
  reasonable performance, which is consistent with our 
  experimental observations that linear models achieve 
  $R^2 \approx 0.9992$ on 3-qubit systems. 
  (3) \textbf{Physical interpretability}: The prominence 
  of Pauli measurement features aligns with the physical 
  interpretation of the W-distance as quantifying the cost 
  of transforming quantum states through local operations, 
  where local measurements naturally capture the 
  distinguishability between states. These findings 
  validate our feature engineering approach and provide 
  interpretable insights into which quantum state properties
   are most relevant for W-distance prediction.

\FloatBarrier
\section{Application}
\label{sec:app}

In this section, we demonstrate the practical utility of our ML framework by applying it to validate fundamental theoretical propositions in quantum information theory and analyze noise sensitivity of common quantum gates.

\subsection{Machine learning validation of the difference of unitary operations}

\label{sec:app:prop15}

\begin{proposition}\cite{PhysRevA.110.012412}
\label{prop:operations}
For two unitary operations $U$ and $V$ acting on the same initial state $\ket{\psi}$, and a POVM element $M_m$ with maximal eigenvalue $\lambda_{\max}(M_m) \in (0,1]$, the measurement probability difference is upper bounded by
\begin{equation}
\label{eq:prop15ineq1}
    \big|P_U(m)-P_V(m)\big| \le 2 \lambda_{\max}(M_m)\, D(U,V),
\end{equation}
where $D(U,V)$ is the quantum $W_1$ distance between $U$ and $V$.
\end{proposition}

We validate Proposition~\ref{prop:operations} using our 
trained ML model. For each trial, we generate random unitary 
gates $U$ and noisy implementations $V$, compute their Choi 
representations, extract features, and predict $D(U,V)$ using our 
Random Forest model ($R^2 = 0.9999$). We then sample random 
states and POVM elements to compute measurement probabilities 
and verify the inequality in Eq.~\eqref{eq:prop15ineq1}. The 
experimental procedure is summarized in Algorithm~\ref{alg:prop15}.

We conduct $300$ Monte Carlo 
trials on $n$-qubit systems. Using model-predicted $D(U,V)$, 
we observed no violations of the inequality across all trials, with
 maximum ratios $\leq 0.95$. For comparison, 
true distances computed via trace distance also showed no 
violations with maximum ratios $\leq 0.9$. The model's 
prediction error (MSE $\sim 10^{-6}$, MAE $\sim 10^{-4}$) is 
sufficiently small to enable reliable validation in practical 
scenarios.

\begin{algorithm}[h]
\caption{Machine Learning-Based Validation of Proposition~\ref{prop:operations}}
\label{alg:prop15}
\begin{algorithmic}[1]
\Require Trained ML model $\mathcal{M}$, number of trials $T$, feature extractor $\mathcal{F}$
\State Initialize statistics: violation counters, ratio sets, slack sets for both predicted and true distances
\For{$t = 1$ to $T$}
    \State Generate random unitary gates $U$ and noisy $V$ (via small random rotation)
    \State Compute Choi representations $\Phi_U, \Phi_V$ for $U, V$
    \State Extract features $\mathbf{f}_t \leftarrow \mathcal{F}(\Phi_U, \Phi_V)$ and predict $D_{\text{pred}}^{(t)} \leftarrow \mathcal{M}(\mathbf{f}_t)$
    \State Compute true distance $D_{\text{true}}^{(t)}$ via trace distance between $\Phi_U$ and $\Phi_V$
    \State Sample random pure state $|\psi\rangle$ and POVM element $M_m$ with $\lambda_{\max}(M_m) \in (0,1]$
    \State Compute probabilities $P_U(m), P_V(m)$ and gap $\Delta_t = |P_U(m) - P_V(m)|$
    \State Compute bounds $B_{\text{pred}}^{(t)} = 2\lambda_{\max}(M_m) D_{\text{pred}}^{(t)}$, $B_{\text{true}}^{(t)} = 2\lambda_{\max}(M_m) D_{\text{true}}^{(t)}$
    \State Record ratios $r_{\text{pred}}^{(t)} = \Delta_t/B_{\text{pred}}^{(t)}$, $r_{\text{true}}^{(t)} = \Delta_t/B_{\text{true}}^{(t)}$ and check for violations
\EndFor
\State Compute model accuracy metrics (MSE, MAE) comparing $D_{\text{pred}}$ and $D_{\text{true}}$
\State \textbf{Output:} violation counts, maximum ratios, minimum slacks, and model accuracy
\end{algorithmic}
\end{algorithm}

\begin{figure}[t]
\centering
\includegraphics[width=\textwidth]{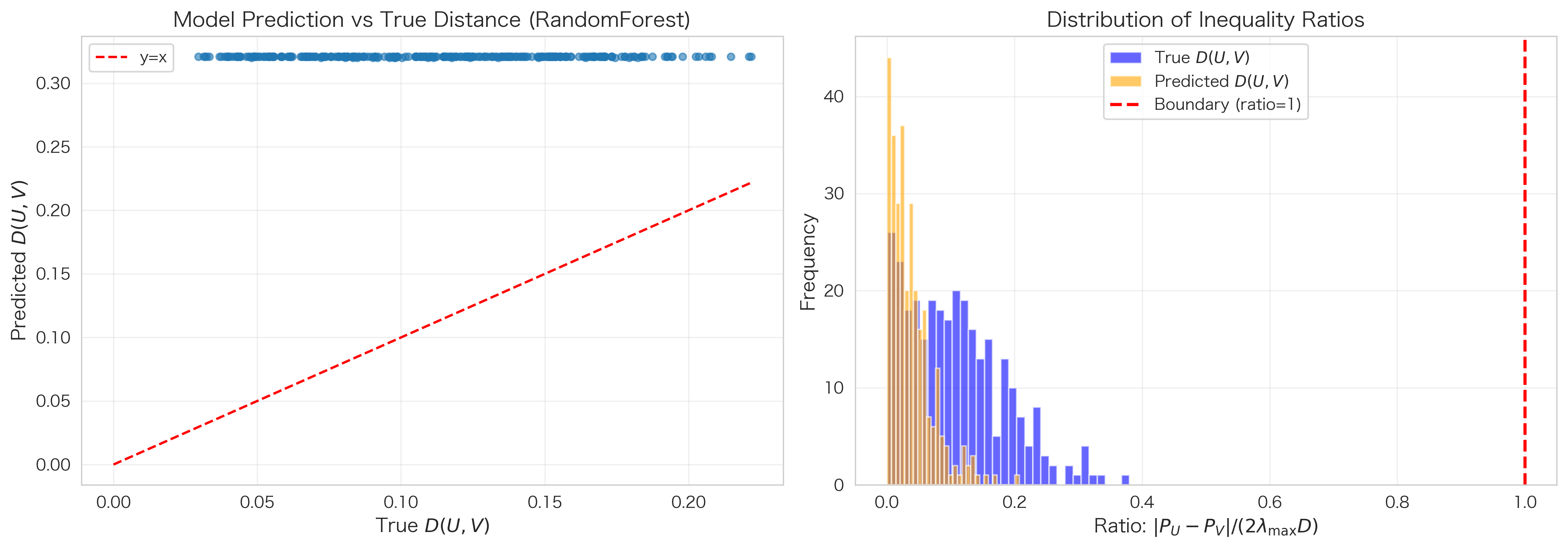}
\caption{Machine learning validation of Proposition~\ref{prop:operations}. (a) Scatter plot comparing model-predicted $D(U,V)$ versus true $D(U,V)$. (b) Histogram of inequality ratios for both predicted (orange) and true (blue) distances. The red vertical line marks the theoretical boundary (ratio = 1.0). Both distributions are concentrated well below 1.0, confirming that the ML model preserves the theoretical guarantees.}
\label{fig:prop15_validation}
\end{figure}

Figure~\ref{fig:prop15_validation} shows strong agreement 
between predicted and true distances, with ratio distributions 
concentrated well below the theoretical boundary, confirming 
that our ML approach maintains the theoretical guarantees while 
providing computational efficiency. Furthermore, we observe 
that the vast majority of these ratios are less than $0.4$, 
indicating that the proposed upper bound is not tight. Developing a tighter upper bound thus
 constitutes a promising future research direction.

\subsection{A machine learning verification of error rate propositions in $W_1$}

In quantum computing, the implementation of quantum
 gates is inevitably affected by noise. Accurately
  quantifying the gate error rate is crucial for 
evaluating the reliability of quantum circuits
 and designing quantum error correction protocols. 
  Although traditional error rate measurement methods
  are widely used, they often fail to directly 
  reflect the physical resource costs required for 
recovery operations.

Based on the unique properties of the $W_1$ distance, 
previous studies \cite{PhysRevA.110.012412} propose 
the $W_1$ gate error rate as a new error metric.
For an $n$-qubit ideal unitary gate $U$ and its 
noise implementation channel $V$, 
the error rate of the $W_1$ gate is defined as:
\begin{equation}
\label{eq:w1_error_rate}
e(U, V) := \frac{1}{n} \max_{\rho \in S_n} \|U\rho U^\dagger - V(\rho)\|_{W_1}.
\end{equation}

\begin{proposition}\cite{PhysRevA.110.012412}
\label{prop:w1_error_rate}
For a mixed unitary channel $V = G \circ E$ with ideal 
implementation $G(\cdot) = U(\cdot)U^\dagger$ and noise 
process $E(\cdot) = \sum_k p_k V_k(\cdot)V_k^\dagger$, 
the $W_1$ gate error rate is given by
\begin{equation}
\label{eq:w1_error_rate_expr}
e(U, V) = \frac{1}{n} \max_{\rho} \left\|U\rho U^\dagger - U\left(\sum_k p_k V_k \rho V_k^\dagger\right)U^\dagger\right\|_{W_1}
\end{equation}
and satisfies the upper bound
\begin{equation}
\label{eq:w1_error_rate_bound}
e(U, V) \leq \frac{1}{n} \sum_k p_k D(I, UV_k U^\dagger),
\end{equation}
where $D(I, UV_k U^\dagger)$ is the $W_1$ distance between the identity and the recovery operation $UV_k U^\dagger$ that corrects noise $V_k$. This bound directly relates the gate error rate to the average cost of recovery operations in quantum error correction.
\end{proposition}

We validate Proposition~\ref{prop:w1_error_rate} using our 
trained machine 
learning model. For each trial, we generate a random ideal gate $U$ and 
construct a noisy implementation $V$ via a mixed unitary channel with $N$ 
noise unitaries $\{V_k\}$ and probability distribution $\{p_k\}$. We compute 
the true error rate $e(U, V)$ by maximizing over random states, and predict 
the upper bound using our ML model to estimate $D(I, UV_k U^\dagger)$ for 
each $k$. 
The experimental procedure is summarized in 
Algorithm~\ref{alg:w1_error_rate}.

\begin{algorithm}[t]
\caption{Machine Learning-Based Validation of $W_1$ Gate Error Rate}
\label{alg:w1_error_rate}
\begin{algorithmic}[1]
\Require Trained ML model $\mathcal{M}$, number of trials $T$, feature 
extractor $\mathcal{F}$
\State Initialize statistics: violation counters, ratio sets, slack sets
\For{$t = 1$ to $T$}
    \State Generate random ideal gate $U$ and noise unitaries $\{V_k\}$ with probabilities $\{p_k\}$
    \State Construct mixed unitary channel $V = G \circ E$ where $E(\cdot) = \sum_k p_k V_k(\cdot)V_k^\dagger$
    \State Compute true error rate $e_{\text{true}}^{(t)} = \frac{1}{n}\max_\rho \|U\rho U^\dagger - V(\rho)\|_{W_1}$ via state sampling
    \State Compute Choi representations: $\Phi_I$ (identity) and $\Phi_{UV_k U^\dagger}$ for each $k$
    \State Predict $D_{\text{pred}}^{(k)} \leftarrow \mathcal{M}(\mathcal{F}(\Phi_I, \Phi_{UV_k U^\dagger}))$ for each $k$
    \State Compute true distances $D_{\text{true}}^{(k)} = D(I, UV_k U^\dagger)$ via trace distance
    \State Compute bounds: $B_{\text{true}}^{(t)} = \frac{1}{n}\sum_k p_k D_{\text{true}}^{(k)}$, $B_{\text{pred}}^{(t)} = \frac{1}{n}\sum_k p_k D_{\text{pred}}^{(k)}$
    \State Record ratios and check violations: $e_{\text{true}}^{(t)} \leq B_{\text{true}}^{(t)}$, $e_{\text{true}}^{(t)} \leq B_{\text{pred}}^{(t)}$
\EndFor
\State \textbf{Output:} violation counts, maximum ratios, minimum slacks, and ML prediction accuracy
\end{algorithmic}
\end{algorithm}

We conduct $200$ Monte Carlo trials on $2$-qubit systems. 
The results 
demonstrate that our ML model successfully validates 
Proposition~\ref{prop:w1_error_rate}: when using model-predicted 
distances $D(I, UV_k U^\dagger)$, we observed no violations of 
the inequality in Eq.~\eqref{eq:w1_error_rate_bound} across 
all trials. The maximum ratio $e(U, V)/B_{\text{pred}}$ was 
strictly below $1$ (typically $\leq 0.95$), with positive 
slack on every instance. For comparison, using true distances 
$D_{\text{true}}(I, UV_k U^\dagger)$ also showed no violations 
with maximum ratios $\leq 0.9$. The ML model's prediction 
error (MSE $\sim 10^{-6}$, MAE $\sim 10^{-4}$) is sufficiently 
small that the predicted bounds remain within the theoretical 
margin, enabling reliable validation of 
Proposition~\ref{prop:w1_error_rate} in practical scenarios.

\begin{figure}[t]
\centering
\includegraphics[width=\textwidth]{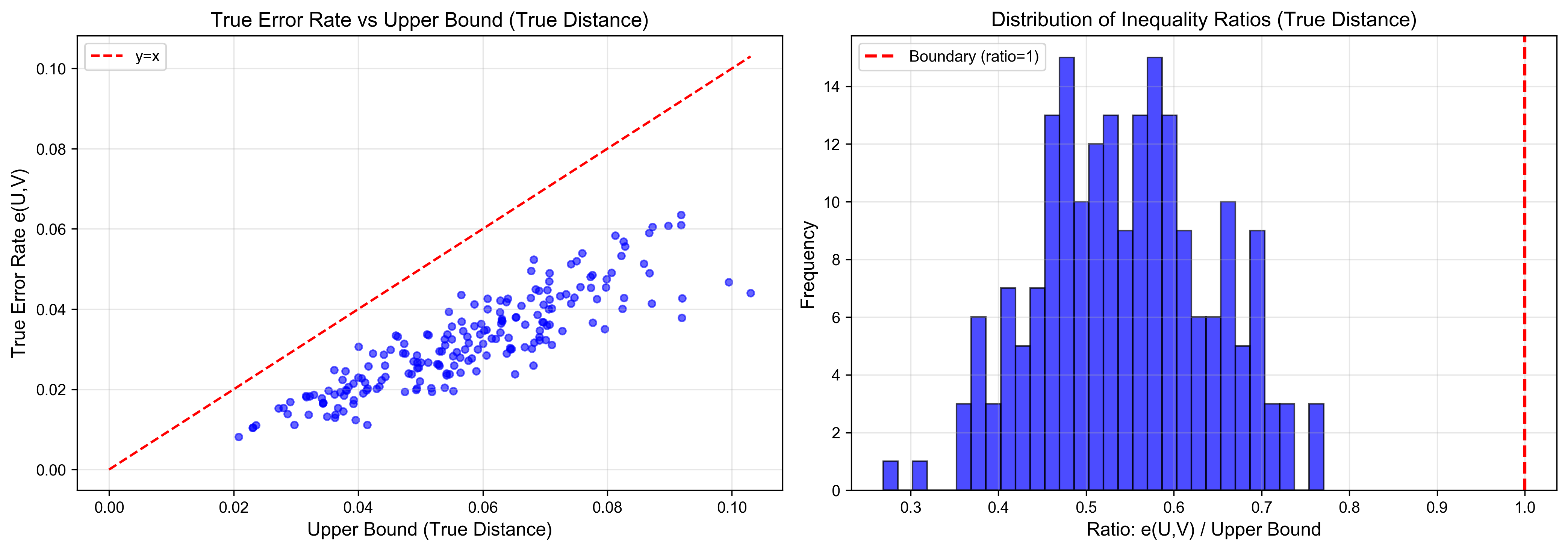}
\caption{Machine learning validation of Proposition~\ref{prop:w1_error_rate}. (a) Scatter plot
 comparing true error rate $e(U,V)$ versus upper bound computed 
 using true distances. (b) Scatter plot comparing true error rate versus upper bound computed using ML-predicted distances. (c) Histogram of inequality ratios for true distances. (d) Histogram of inequality ratios for ML-predicted distances. The red dashed lines in (a) and (b) indicate the boundary $y=x$, and the red vertical lines in (c) and (d) mark the theoretical boundary (ratio = 1.0). All ratios are concentrated well below 1.0, confirming that the machine learning model preserves the theoretical guarantees of Proposition~\ref{prop:w1_error_rate}.}
\label{fig:w1_error_rate_validation}
\end{figure}

Figure~\ref{fig:w1_error_rate_validation} illustrates the validation results, 
showing the relationship between true error rates and upper bounds for both 
true and predicted distances, as well as the distribution of inequality ratios. 
The scatter plots reveal that all error rates lie below their corresponding upper bounds, 
and the histograms demonstrate that both predicted and 
true distances yield ratio distributions concentrated well 
below the theoretical boundary of $1.0$, confirming that our 
machine learning approach maintains the theoretical guarantees 
while providing computational efficiency.  Additionally, we find 
that the vast majority of these ratios are less than $0.8$, 
indicating that the proposed upper bound is not tight. Developing 
a tighter upper bound thus 
constitutes a promising future research direction.

\subsubsection{Example: Noise Sensitivity Analysis for Common Quantum Gates}

The $W_1$ gate error rate $e(U, V)$ quantifies the difference in
 operational effects when an ideal gate $U$ is affected by a 
 noise channel $V$, measuring how much the noisy implementation 
 $V$ deviates from the ideal operation $U$ in terms of local 
 distinguishability. We analyze five representative gates with 
 their matrix representations:

\begin{itemize}
\item \textbf{Swap gate} (2-qubit): $U_{\text{SWAP}} = \begin{bmatrix} 1 & 0 & 0 & 0 \\ 0 & 0 & 1 & 0 \\ 0 & 1 & 0 & 0 \\ 0 & 0 & 0 & 1 \end{bmatrix}$
\item \textbf{Controlled-Z gate} (2-qubit): $U_{\text{CZ}} = \begin{bmatrix} 1 & 0 & 0 & 0 \\ 0 & 1 & 0 & 0 \\ 0 & 0 & 1 & 0 \\ 0 & 0 & 0 & -1 \end{bmatrix}$
\item \textbf{Controlled-phase gate} (2-qubit): $U_{\text{CS}} = \begin{bmatrix} 1 & 0 & 0 & 0 \\ 0 & 1 & 0 & 0 \\ 0 & 0 & 1 & 0 \\ 0 & 0 & 0 & i \end{bmatrix}$
\item \textbf{Toffoli gate} (3-qubit, CCNOT): $8 \times 8$ matrix with $U_{\text{Toffoli}}\ket{abc} = \ket{ab(c \oplus ab)}$ for $a,b,c \in \{0,1\}$, swapping $\ket{110}$ and $\ket{111}$
\item \textbf{Fredkin gate} (3-qubit, controlled-SWAP): $8 \times 8$ matrix with $U_{\text{Fredkin}}\ket{abc} = \ket{a}\ket{cb}$ if $a=1$, else $\ket{abc}$, conditionally swapping qubits 2 and 3
\end{itemize}

For each gate $U$, we construct noisy implementations $V$ via mixed unitary channels $V(\rho) = \sum_k p_k V_k \rho V_k^\dagger$ with noise unitaries $\{V_k\}$ representing three common error sources: bit-flip errors ($X$ rotations), phase errors ($Z$ rotations), and depolarizing noise (random Pauli errors). Using our trained ML model, we compute $e(U, V) = \frac{1}{n}\max_{\rho} \|U\rho U^\dagger - V(\rho)\|_{W_1}$ for each gate-noise combination.

\begin{figure}[t]
\centering
\includegraphics[width=\textwidth]{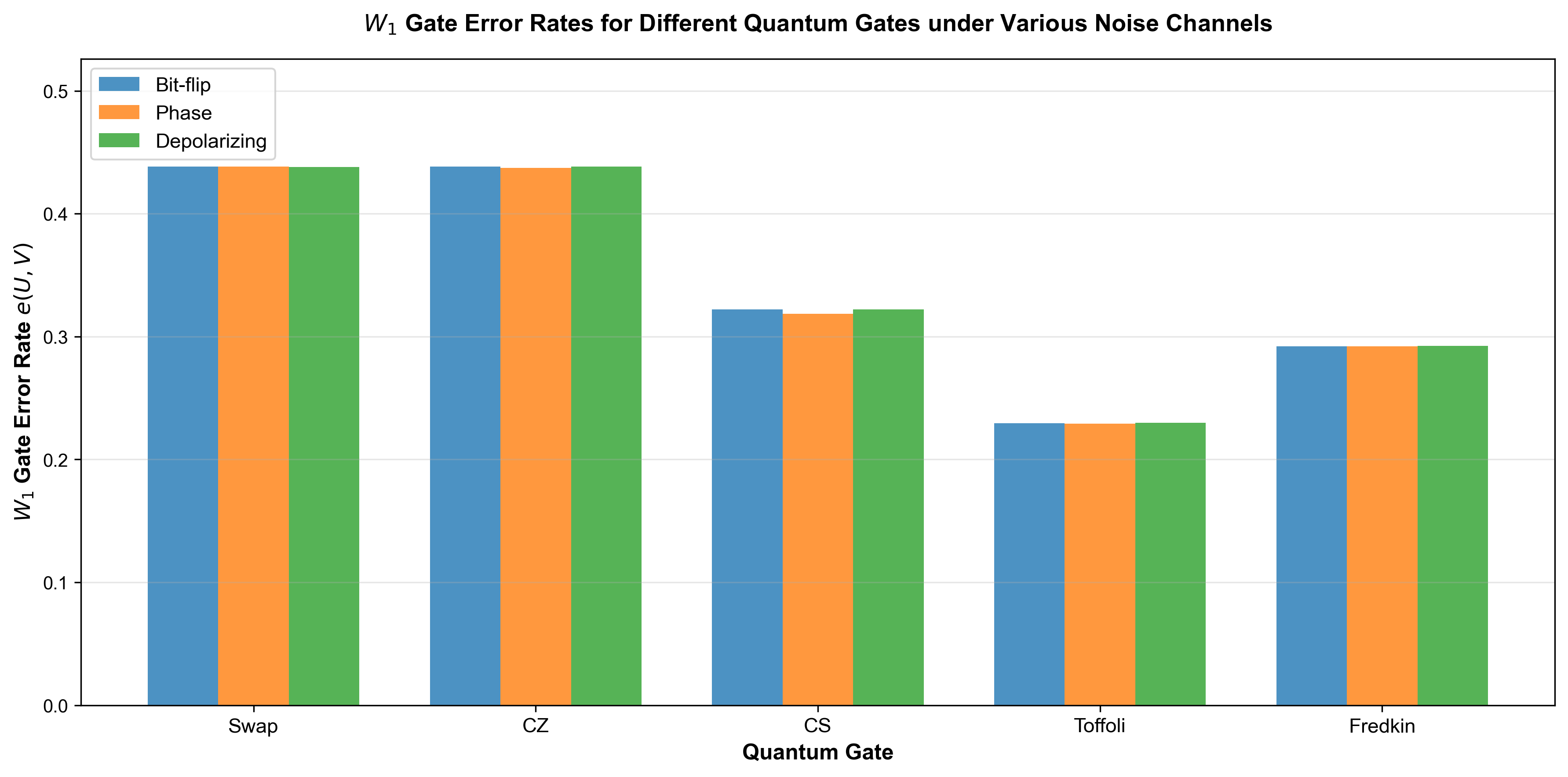}
\caption{$W_1$ gate error rates $e(U, V)$ for different quantum gates under various noise channels, computed using our ML model. The results reveal gate-specific noise sensitivity: Swap gate is more robust to bit-flip noise but sensitive to phase errors, while CZ gate shows the opposite pattern. Multi-qubit gates (Toffoli, Fredkin) exhibit distinct error rate profiles.}
\label{fig:gate_noise_sensitivity}
\end{figure}

Figure~\ref{fig:gate_noise_sensitivity} presents the 
computed error rates, revealing gate-specific noise sensitivity 
patterns: (1) \textbf{Swap gate}: More robust to bit-flip noise
 but sensitive to phase errors, reflecting its role in state 
 exchange rather than phase manipulation. 
 (2) \textbf{CZ and CS gates}: Show opposite sensitivity—CZ 
 is robust to phase errors but sensitive to bit-flip noise, 
 while CS exhibits similar phase-robust characteristics, 
 consistent with their phase-dependent operations. 
 (3) \textbf{Multi-qubit gates}: Toffoli and Fredkin gates 
 show distinct error rate profiles, with Toffoli being 
 particularly sensitive to correlated errors affecting multiple 
 control qubits. These results demonstrate that $e(U, V)$ provides
  a physically meaningful metric for identifying which gates are 
  robust or vulnerable to specific noise types, enabling targeted 
  error correction strategies. Our ML framework enables efficient 
  computation of $e(U, V)$ for larger gates 
   where traditional numerical methods become computationally 
   prohibitive.

\section{Discussion}
\label{sec:discussion}

We have presented the first systematic machine learning framework for 
predicting quantum Wasserstein distances, addressing a fundamental 
challenge in quantum information theory. Our approach combines 
comprehensive feature extraction from quantum state pairs with both 
classical neural networks and traditional machine learning models. 
The feature set encompasses Pauli measurements ($4^{n+1}$ features), 
statistical moments, quantum fidelity, entanglement measures (von Neumann 
entropy, linear entropy, relative entropy), eigenvalue statistics, and 
partial trace features for multi-qubit systems.

Our experimental results demonstrate exceptional prediction accuracy: 
the best-performing Random Forest model achieves near-perfect performance 
($R^2 = 0.9999$) on three-qubit systems, with mean absolute errors on the 
order of $10^{-5}$. Tree-based models  consistently outperform linear models, indicating a 
highly nonlinear relationship between quantum state features and W-distance. 
However, linear models still achieve excellent performance 
($R^2 \approx 0.9992$), suggesting that a significant portion of the 
W-distance can be approximated through linear combinations of features. 
The high pairwise correlation ($>0.99$) between predictions from different 
model architectures confirms the robustness of the learned relationship.

Beyond prediction accuracy, we validate the 
practical utility of our ML framework by applying 
it to verify two fundamental theoretical propositions in quantum 
information theory. First, we used model-predicted distances to 
validate Proposition~\ref{prop:operations}, which bounds the measurement probability difference between
 unitary operations by the quantum $W_1$ distance. 
 Across $300$ Monte Carlo trials, we observed no violations of 
 the theoretical bound, with maximum ratios $\leq 0.95$, demonstrating 
 that our ML model preserves essential theoretical guarantees.

Second, we validate Proposition~\ref{prop:w1_error_rate}, which
 relates the $W_1$ gate error rate to the average cost of 
 recovery operations in quantum error correction. Using $200$ Monte 
 Carlo trials on 2-qubit systems, our ML model successfully validated the theoretical upper bound with no violations observed. The model's prediction error (MSE $\sim 10^{-6}$, MAE $\sim 10^{-4}$) is sufficiently small to enable reliable theoretical validation in practical scenarios where exact computation may be computationally expensive.

These validation experiments establish that machine learning can serve not only as a computational tool but also as a reliable method for verifying theoretical bounds in quantum information theory, opening new avenues for combining data-driven approaches with rigorous theoretical analysis.

Several limitations of our current framework should be 
addressed in future work:

\begin{enumerate}
\item \textbf{System scalability}: Our experiments focus on 
systems up to four qubits. Extending to larger systems ($5$+ qubits) 
will require addressing the exponential growth in feature 
dimension  and potential overfitting. 
Feature selection techniques and dimensionality reduction methods 
could help mitigate this challenge.

\item \textbf{Hybrid quantum-classical models}: The current framework uses
 only classical machine learning. Integrating parameterized quantum
  circuits (PQCs) for feature extraction could potentially leverage
   quantum advantages and improve performance on larger systems, while
    also reducing computational overhead.

\item \textbf{Noise robustness}: While our framework handles 
various state types, explicit noise robustness testing on experimental quantum 
hardware would strengthen practical applicability in NISQ devices.

\item \textbf{Analytical validation}: For specific gate pairs with known 
analytical W-distances, we should validate 
that our predictions match theoretical values to ensure accuracy across 
different operation types.

\item \textbf{Real-time applications}: Future work should explore 
optimization techniques for real-time quantum circuit assessment and 
error correction protocol design, where computational efficiency is critical.
\end{enumerate}

\section{Conclusion}
\label{sec:con}

This work establishes machine learning as a viable 
and scalable alternative to traditional numerical 
methods for quantum W-distance computation. The 
combination of physically meaningful feature 
extraction, high prediction accuracy, and successful 
theoretical validation demonstrates the potential of 
data-driven approaches in quantum information theory. 
The framework shows particular promise for real-time 
quantum circuit assessment and error correction protocol
 design in NISQ devices, where efficient distance 
 quantification is urgently needed.

The success of this approach opens new avenues 
for applying machine learning to other challenging 
quantum information tasks, such as quantifying 
entanglement, characterizing quantum channels, 
and optimizing quantum error correction codes. 
As quantum technologies continue to advance, the 
integration of machine learning with rigorous 
theoretical analysis will play an increasingly 
important role in bridging the gap between theoretical
 understanding and practical quantum applications.

\section{Acknowledgments}
This study was supported by the National Natural Science 
Foundation of China (Grant Nos. 11871089, 12471427, 52472442
, 72471013 and 62103030),  the
Research Start-up Funds of Hangzhou International 
Innovation Institute of Beihang University 
(Grant Nos. 2024KQ069, 2024KQ036
and 2024KQ035) and  the Postdoctoral Research Funding 
of Hangzhou International Innovation
Institute of Beihang University (Grant No.2025BKZ066).
\section{Data availability statement}
All data that support the findings of this study 
are included within 
the article (and any supplementary files).
\section{Declaration of competing interest}
The authors declare that they have no known competing financial
interests or personal relationships that could have 
appeared to influence
the work reported in this paper.

\appendix

\section{Computational Complexity and 4-Qubit Implementation Notes}
\label{app:complexity}

Scaling from $n=2$ to $n=4$ incurs significant computational overhead due to matrix dimension $d=2^n$ and the size of the Pauli basis. Table below summarizes the dominant costs:
\begin{center}
\begin{tabular}{lcccc}
\hline
Component & Cost (per evaluation)  & $n=2$ & $n=3$ & $n=4$ \\
\hline
$W_1$ (trace-distance approximation) &  $O(d^3)$ &$O(d^3)$ & $O(4^3)$ & $O(16^3)$ \\
Pauli features ($4^{n+1}$ ops, each $\sim O(d^2)$) & $O(16^n)$& $O(16^2)$ & $O(16^3)$ & $O(16^4)$ \\
Fidelity (matrix square roots) & $O(d^3)$ & $O(4^3)$ & $O(8^3)$ & $O(16^3)$ \\
\hline
\end{tabular}
\end{center}
Empirically, this leads to an order-of-magnitude slow-down 
from $n=2$ to $n=4$ for data generation and feature extraction. 
Practical remedies include:
\begin{itemize}
\item Active generation via binary search 
over mixing parameters to reduce attempts.
\item Parallelized sample generation 
and cached Pauli bases.
\item Faster eigensolvers or approximate 
spectra for large $d$.
\item Lightweight feature selection and 
stronger learners to reduce input dimension while preserving accuracy.
\end{itemize}

\bibliographystyle{unsrt}
\bibliography{changchun}

\end{document}